\documentclass[runningheads]{llncs}
\usepackage{graphicx}
\usepackage{xcolor}
\usepackage{amsmath}
\usepackage{booktabs}
\usepackage{amsfonts}
\usepackage{chngcntr}
\usepackage{subfigure}
\usepackage{graphicx}
\usepackage[hyperfootnotes]{hyperref}

\begin{document}
\title{Learning Tumor Growth via Follow-Up Volume Prediction for Lung Nodules}
%
\author{Yamin Li\inst{1,2,3,}\thanks{These authors have contributed equally: Y. Li. and J. Yang.} \and Jiancheng Yang\inst{1,2,3,\star} \and Yi Xu \inst{1,2,3,}\thanks{Corresponding author: Yi Xu (xuyi@sjtu.edu.cn).} \and
Jingwei Xu \inst{1,2,3} \and \\Xiaodan Ye \inst{1,4} \and Guangyu Tao \inst{1,4} \and Xueqian Xie \inst{1,5} \and Guixue Liu \inst{1,5}}

\authorrunning{Y. Li et al.}

\institute{Shanghai Jiao Tong University, Shanghai, China \and
Shanghai Institute for Advanced Communication and Data Science \and
MoE Key Lab of Artificial Intelligence, AI Institute
Shanghai Jiao Tong University \and
Shanghai Chest Hospital, Shanghai, China \and
Shanghai General Hospital, Shanghai, China\\
\email{\{yaminli, jekyll4168, xuyi\}@sjtu.edu.cn}}
\maketitle              

\begin{abstract}
Follow-up serves an important role in the management of pulmonary nodules for lung cancer. Imaging diagnostic guidelines with expert consensus have been made to help radiologists make clinical decision for each patient. However, tumor growth is such a complicated process that it is difficult to stratify high-risk nodules from low-risk ones based on morphologic characteristics. On the other hand, recent deep learning studies using convolutional neural networks (CNNs) to predict the malignancy score of nodules, only provides clinicians with black-box predictions. To this end, we propose a unified framework, named Nodule Follow-Up Prediction Network (\textit{NoFoNet}), which predicts the growth of pulmonary nodules with high-quality visual appearances and accurate quantitative 
results, given any time interval from baseline observations. It is achieved by predicting future displacement field of each voxel with a WarpNet. A TextureNet is further developed to refine textural details of WarpNet outputs. We also introduce techniques including Temporal Encoding Module and Warp Segmentation Loss to encourage time-aware and shape-aware representation learning. We build an in-house follow-up dataset from two medical centers to validate the effectiveness of the proposed method. \textit{NoFoNet}~significantly outperforms direct prediction by a U-Net in terms of visual quality; more importantly, it demonstrates accurate differentiating performance between high- and low-risk nodules. Our promising results suggest the potentials in computer aided intervention for lung nodule management.


\end{abstract}

\section{Introduction}

Pulmonary nodule management strategy influences the cost-effectiveness of a lung cancer screening program \cite{Cressman2014ResourceUA}. It remains difficult to differentiate high-risk nodules from low-risk ones based on morphologic characteristics \cite{Pinsky2013NationalLS}. In order to help radiologists and clinicians to make precise clinical decision for each patient, researchers have made several categorical management recommendation and scoring systems according to morphology, diameters or volume in recent years, \textit{e.g.}, NCCN~\cite{Wood2015NationalCC}, Fleischner~\cite{MacMahon2017GuidelinesFM}, Lung-RADS~\cite{Pinsky2015PerformanceOL}. However, tumor growth is such a complicated progress that more advanced strategies are worth exploring to facilitate precision medicine. Emerging deep learning technology suggests a potential alternative to develop end-to-end lung nodule management system in a data-driven fashion. Although numerous studies have explored end-to-end approaches to predict malignancy scores \cite{Hussein2017RiskSO,Xie2017TransferableME,yang2019probabilistic} or categories \cite{yang2020relational,zhao2019toward,zhao20183d} of lung nodules, while only a few studies \cite{Ardila2019EndtoendLC,Huang2019PredictionOL} address the lung nodule follow-up problem. Nevertheless, these studies only provide black-box predictions without intuitive explanations. There is also study \cite{petersen2019deep} on predicting tumor growth with a model-free appearance modeling approach using a probabilistic U-Net \cite{ronneberger2015u}, however it could not provide any quantitative assessment on the risk of tumors.

In this study, we aim at a unified approach to predict growth of lung nodules, with both high-quality visual appearances and accurate quantitative results. The core of our approach is based on a WarpNet, predicting displacement field $\mathbf{u}$ (or motion \cite{jin2017predicting}) on a future volume from a baseline volume. With the field  $\mathbf{u}$, we could obtain not only the predicted \textbf{visual appearance} of the future volume by warpping the baseline, but also the feature segmentation mask from the baseline mask, which could be used for \textbf{quantitative assessment} of tumor growth. This approach is inspired from VoxelMorph \cite{ronneberger2015u}, where the displacement field for registration is conditional on both the baseline and future volumes; instead, our predictive displacement field is conditional only on the baseline volume and could be dynamically estimated. Moreover, a TextureNet is designed to refine textural details of the outputs from WarpNet. We introduce techniques including Temporal Encoding Module and Warp Segmentation Loss to encourage time-aware and shape-aware representation learning. The whole network, named Nodule Follow-Up Prediction Network (\textit{NoFoNet}), establishes a unified framework to produce both high-quality visual appearances and accurate quantitative assessment for lung nodule follow-up. Our in-house follow-up dataset from two medical centers validates the effectiveness of \textit{NoFoNet}.

\section{Materials and Methods}

\subsection{Task Formalization and Dataset}

We aim at a unified framework to predict future volume of a lung nodule, given any time interval and a baseline volume. An in-house dataset is collected, containing 622 LDCT scans from 246 patients (114 males and 132 females) with a total of 315 long-standing pulmonary nodules. Each patient has at least two time points of thin layer LDCT (slice thickness $\leq$ 1.25mm), with the time interval of 30-1351, 136 days (min-max, median). We select nodules at every two time points as a sample (for example if a nodule has 3 follow-up scans at time points $t_1t_2t_3$, we choose time points $t_1\&t_2$, $t_1\&t_3$, $t_2\&t_3$ as 3 samples), resulting in 731 pairs. The age of the patients at first examination is 23-97, 62 years. The segmentation VOI of each selected nodule (diameter from 3mm to 30mm) is delineated by an expert radiologist and checked by another. 


We pre-process the data as follows~\cite{yang2019probabilistic}: CT scans are resampled isotropically into $1mm\times1mm\times1mm$. The voxel intensity is normalized to $\left[-1, 1\right)$ from the Hounsfield unit (HU), using the mapping function $I = \lfloor \frac{I_{HU}+1024}{400+1024} \times 255\rfloor / 128 - 1$. Each data sample is a cubic volume image with the size of $48\times48\times48$, which covers the size of all nodules in our study.

\begin{figure}[tb]
\includegraphics[width=\textwidth]{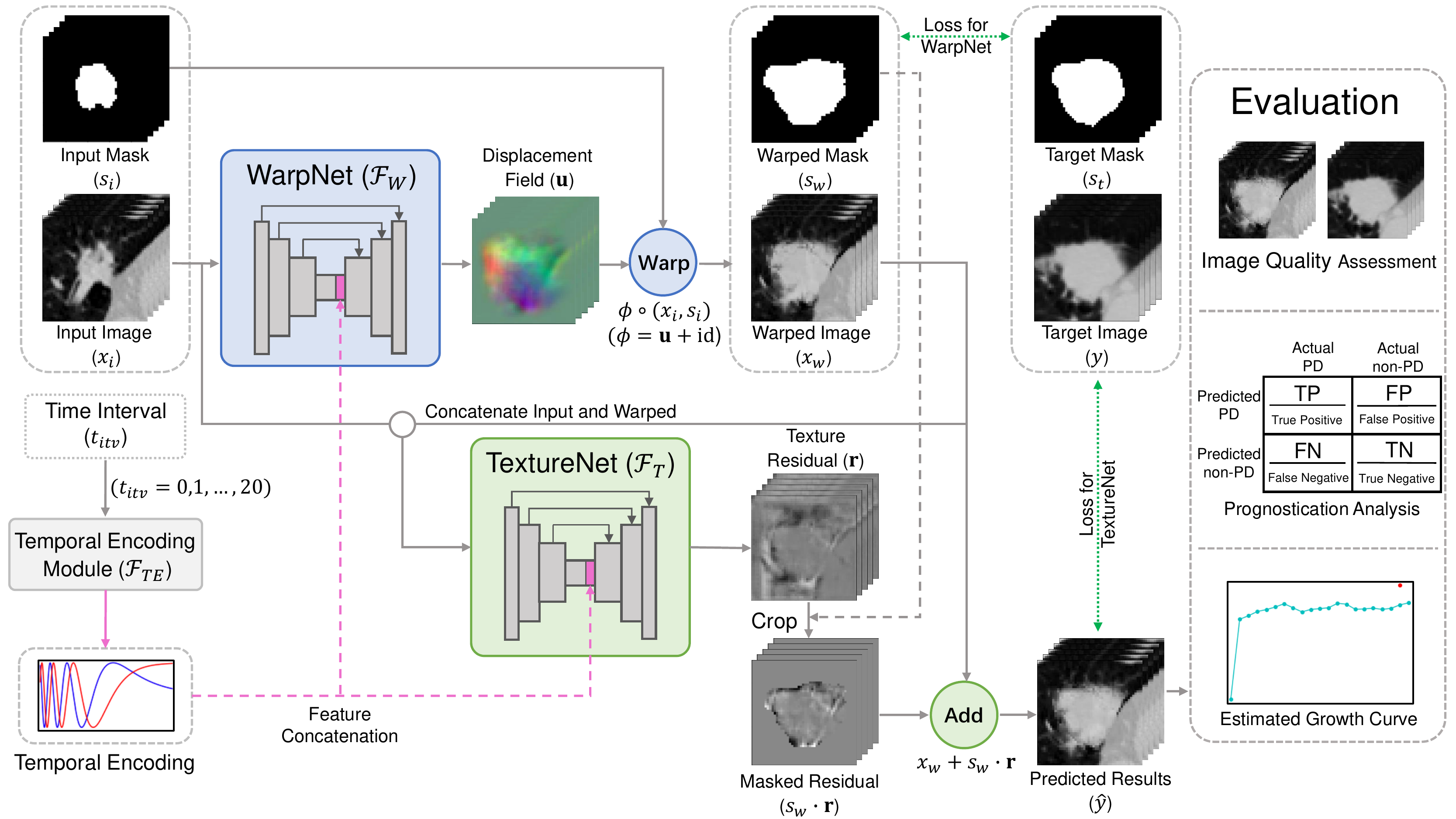}
\caption{Overview of the proposed \textit{NoFoNet} architecture. \textit{NoFoNet} consists of a WarpNet and a TextureNet, each with a Temporal Encoding Module (TEM), which encodes follow-up time interval into the nodule representation. WarpNet and TextureNet are two 3D CNNs based on U-Net~\cite{ronneberger2015u}, modeling the spatial and texture transformation for nodule growth respectively. During evaluation, we perform the image quality assessment, prognostication analysis and growth estimation. Note PD means progressive disease, i.e., significant growth of nodule in our study. The red point in the growth curve represents actual volume of the nodule in the future time.} \label{fig1}
\end{figure}

\subsection{\textit{NoFoNet}: Nodule Follow-Up Prediction Network}

To model the growth of nodules, we develop a Nodule Follow-Up Prediction Network (\textit{NoFoNet}, see Fig.~\ref{fig1}) consisting of a WarpNet $\mathcal{F}_W$ and a TextureNet $\mathcal{F}_T$ for spatial and texture (intensity) transformations~\cite{zhao2019data} respectively, where an integrated temporal encoding module (TEM) $\mathcal{F}_{TE}$ is addressed to encode different follow-up time interval information into the lesion representation. As we will show later, the WarpNet and TextureNet are able to model the shape and texture variation of nodule growth well.

Given a pair of follow-up input and target images\footnote[1]{If no otherwise specified, image mentioned here and later in this article refers to $48\times48\times48$ cubic volume image with a nodule in the center.} with time interval $t_{itv}$, each of which has corresponding nodule segmentation map $\{x_i, s_i\}$ and $\{y, s_t\}$, the WarpNet $\mathcal{F}_W$ with parameter $\theta_{w}$ first predicts a smooth voxel-wise displacement field $\mathbf{u} = \mathcal{F}_W(x_i,\mathcal{F}_{TE}(t_{itv});\theta_{w})$ for spatial transformation. Following the registration literature~\cite{balakrishnan2019voxelmorph}, we have the warp function $\phi = \mathbf{u} + \text{id}$, where id is identity function. We apply the warp function $\phi$ to $x_i$ to get the warped image $x_w$, and denote this as $x_w = \phi \circ x_i$. Similarly, the warped segmentation map $s_w = \phi \circ s_i$. The TextureNet $\mathcal{F}_T$ with parameter $\theta_{t}$ takes the concatenation of $x_i$ and $x_w$ as inputs and generates a voxel-wise residual $\mathbf{r} = \mathcal{F}_T(x_i,x_w,\mathcal{F}_{TE}(t_{itv});\theta_{t})$. Then we get the results $\hat{y} = x_w + s_w \cdot \mathbf{r}$, where $s_w \cdot \mathbf{r}$ denotes the residual cropped by warped segmentation. The overall formulation of our \textit{NoFoNet} is as follows:

\begin{equation}
\label{eq1}
\begin{aligned}
&\mathbf{u} = \mathcal{F}_W(x_i,\mathcal{F}_{TE}(t_{itv});\theta_{w}), \quad \phi = \mathbf{u} + \text{id}, \quad x_w = \phi \circ x_i, \quad s_w = \phi \circ s_i; \\
&\mathbf{r} = \mathcal{F}_T(x_i,x_w,\mathcal{F}_{TE}(t_{itv});\theta_{t}), \quad \hat{y} = x_w + s_w \cdot \mathbf{r}.
\end{aligned}
\end{equation}

\subsection{Temporal Encoding Module (TEM)}
Since the time interval between two follow-up scans can be rather different, inspired by positional encoding~\cite{vaswani2017attention} we develop a Temporal Encoding Module (TEM) to embed time interval information into the prediction model. Due to the limitation of dataset size, we discretize the interval using time mapping function $t_{itv} = \lceil t_{day} / 30 \rceil$ with an upper cut-off value 20, for most of the intervals are less than 600 days. Sine and cosine functions with different frequencies are used in the TEM to encode temporal information:
\begin{equation}
\label{eq2}
\begin{aligned}
&\mathcal{F}_{TE}(t_{itv},2i)  = \sin(t_{itv} / 100^{2i / d_{fm}}), \\
&\mathcal{F}_{TE}(t_{itv},2i+1) = \cos(t_{itv} / 100^{2i / d_{fm}}),
\end{aligned}
\end{equation}
where $t_{itv}$ is the discretized time interval, $d_{fm}$ is the total number of channels of the encoded feature map and $i$ is the dimension. That is, the even/odd dimensions of the temporal encoding are generated by sin/cos function with different wavelengths ($2\pi$ to $100\times2\pi$), which makes the relative time information encoded in a redundant way. Besides, the value range of the encoding result is within a certain numerical interval due to the boundedness of sinusoid. These two points ensure that the temporal encoding method can generate a more meaningful high-dimensional representation space. 



\subsection{WarpNet for Spatial Transformation}
As the core of our method, WarpNet predicts a displacement field $\mathbf{u}$ to model the shape variation of nodule growth, which is similar to the motion prediction in video tasks~\cite{jin2017predicting,luc2017predicting,xu2018video}. The architecture of WarpNet is based on a CNN similar to U-Net~\cite{ronneberger2015u} with skip connections, and the temporal encoding from TEM is connected to the bottom of WarpNet. 

The loss function for training WarpNet $\theta_w$ has four terms: similarity loss $\mathcal{L}_{sim}$ between warped images $x_w$ and target images $y$, segmentation loss $\mathcal{L}_{seg}$ between warped segmentation maps $s_i$ and target maps $s_t$, smoothness loss $\mathcal{L}_{smooth}$ for the deformation field and regularization loss $\mathcal{L}_{reg}$ for the output of WarpNet when $t_{itv} = 0$. In summary, the learning of WarpNet is formulated as:
\begin{equation}
\label{eq3}
\hat{\theta_w} = \mathop{\text{argmin}}_{\theta_w}\{\mathcal{L}_{sim}(x_w, y) + \lambda_1\mathcal{L}_{seg}(s_w, s_t) + \lambda_2\mathcal{L}_{smooth}(\mathbf{u}) + \lambda_3\mathcal{L}_{reg}(\phi_0)\}
\end{equation}
with weights $\lambda_1, \lambda_2, \lambda_3 > 0$, where $\phi_0$ is the predicted spatial warp function when time interval $t_{itv} = 0$. All loss functions are designed as follows:

\textit{a) similarity loss and regularization loss:} In our experiments we find that for spatial transformation normalized cross correlation (NCC) loss leads to more reasonable and robust results than MSE loss. The NCC loss between warped image $x_w$ / target image $y$ and the regularization loss for $\phi_0$ is defined as:

\begin{equation}
\label{eq4}
\begin{aligned}
&\mathcal{L}_{sim}(x_w, y) = 1 - NCC(x_w, y) = 1 - NCC(\phi \circ x_i, y), \\
&\mathcal{L}_{reg}(\phi_0) = \mathcal{L}_{sim}(\phi_0 \circ x_i, x_i) + \mathcal{L}_{sim}(\phi_0 \circ y, y).
\end{aligned}
\end{equation}

\textit{b) segmentation loss:} We use Dice loss to constrain the similarity between warped segmentation mask $s_w$ and target mask $s_t$:

\begin{equation}
\label{eq5}
\mathcal{L}_{seg}(s_w, s_t) = 1 - \frac{2\cdot\sum_{p \in \Omega} s_w(p) s_t(p)}{\sum_{p \in \Omega} s_w(p) + \sum_{p \in \Omega} s_t(p)}.
\end{equation}


\textit{c) smoothness loss:} Considering that the contour of nodule changes continuously as it grows, we use a diffusion regularization loss to encourage the smoothness of displacement field $\mathbf{u}$:

\begin{equation}
\label{eq6}
\mathcal{L}_{smooth}(\mathbf{u}) = \frac{1}{|\mathrm{\Omega}|}\sum_{p \in \Omega} ||\nabla\mathbf{u}(p)||^2.
\end{equation}
where finite differences between neighboring voxels are used to approximate the spatial gradients $\nabla\mathbf{u}(p)$ (for x,y,z 3 dimensions).

\subsection{TextureNet for Texture Transformation}
In addition to the shape variation, there is also a texture variation in nodule growth caused by the change of CT value distribution of nodules. So a TextureNet is needed to estimate the residual between warped image $x_w$ and target image $y$. TextureNet follows the architecture of WarpNet. To train TextureNet we need an intensity similarity loss $\mathcal{L}_{sim}^{'}$ between textured images $\hat{y}$ (see Eq.~\ref{eq1}) and target images $y$, and a regularization loss $\mathcal{L}_{reg}^{'}$ for the predicted residual $\mathbf{r_0}$ when $t_{itv} = 0$. So the texture transformation learning is formulated as:
\begin{equation}
\label{eq7}
\hat{\theta_t} = \mathop{\text{argmin}}_{\theta_t}\{\mathcal{L}_{sim}^{'}(\hat{y}, y) + \lambda_1^{'}\mathcal{L}_{reg}^{'}(\mathbf{r_0})\}
\end{equation}
with weight $\lambda_1^{'} > 0$. We choose MSE loss to encourage maximal intensity similarity. The loss functions of TextureNet are defined as:

\begin{equation}
\label{eq8}
\begin{aligned}
&\mathcal{L}_{sim}^{'}(\hat{y}, y) = \frac{1}{|\mathrm{\Omega}|} \sum_{p \in \mathrm{\Omega}}[\hat{y}(p) - y(p)]^2,\\
&\mathcal{L}_{reg}^{'}(\mathbf{r_0}) = \frac{1}{|\mathrm{\Omega}|} \sum_{p \in \mathrm{\Omega}}|\mathbf{r_0}(p)|^2.
\end{aligned}
\end{equation}


\subsection{Implementation Details}

\textit{NoFoNet} can use any CNN architecture for WarpNet and TextureNet, and we use the network design of~\cite{balakrishnan2019voxelmorph} in this work. All of the experiments in this study are implemented on an NVIDA Titan X GPU and an Intel i7-6700 CPU. Our codes are based on Python 3.7.3 and Pytorch-1.2.0~\cite{paszke2017automatic}. We use $\lambda_1 = 0.5, \lambda_2 = 10$ and $\lambda_3=\lambda_1^{'}=1$ for the loss weights in Eq.~\ref{eq3} and Eq.~\ref{eq7}. Online data augmentation methods, including rotation and flipping along a random axis, are applied on the input images. Each part of \textit{NoFoNet} is trained using Adam optimizer~\cite{kingma2014adam} with an initial learning rate of 0.001 for 200 epochs. Specifically, we emphasize the similarity loss inside the segmentation map to put more attention on the nodule.





\section{Experiments}

\subsection{Evaluation Protocol}

\begin{table}[tb]
\caption{Quantitative results of multiple models. We choose U-Net w/ or w/o TEM and WarpNet w/ or w/o segmentation loss as comparisons of \textit{NoFoNet}. The performance is estimated by PSNR, $\mathrm{PSNR}^*$ (PSNR in the nodule parts), dice coefficient between warped/target segmentation maps and sensitivity/specificity/G-mean for PD/non-PD classification. We evaluate the performance of our models on our in-house dataset (see Sec. 2.1) with 5-fold cross validation.}
\label{tab1}
\centering
\begin{tabular*}{\hsize}{@{}@{\extracolsep{\fill}}lcccccc@{}}
\toprule
Method &  PSNR & $\mathrm{PSNR}^*$ & Dice & Sensitivity & Specificity & G-mean \\
\midrule
Baseline (U-Net) &  4.1213 & 29.7490 & - & - & - & - \\
~~+TEM        &  6.0380 & 31.8821 & - & - & - & - \\
\midrule
WarpNet            & 18.0915 & 43.1140 & 0.6301 & 0.7656 & \textbf{0.9083} & 0.8339 \\
~~+Warp Seg Loss   & 18.1952 & 43.2464 & 0.6474 & 0.8594 & 0.8805 & 0.8699 \\
~~+TextureNet      & \textbf{18.2089} & \textbf{43.4904} & \textbf{0.6474} & \textbf{0.8594} & 0.8805 & \textbf{0.8699} \\
\bottomrule
\end{tabular*}
\end{table}

\begin{figure}[tb]
\includegraphics[width=\textwidth]{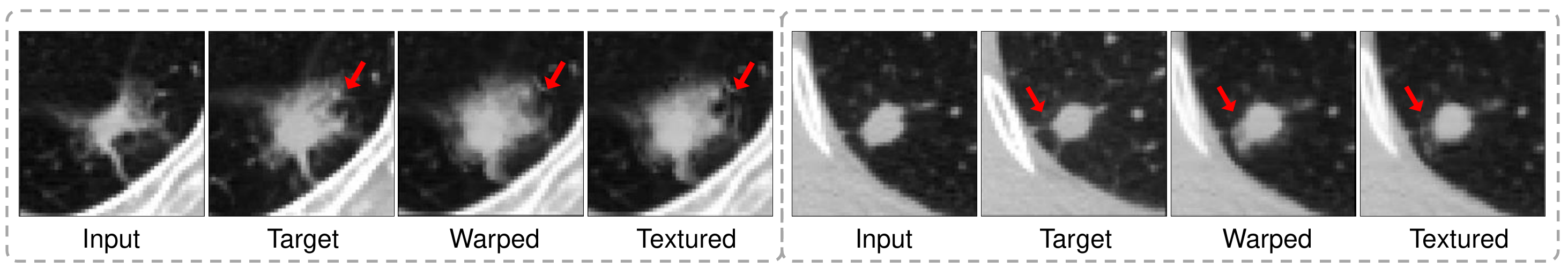}
\caption{Comparison of warped images and textured images for a PD case (left) and a non-PD case (right). Areas where intensity is changed significantly by TextureNet are indicated by red arrows.} \label{fig:warp-texture}
\end{figure}

\begin{figure}[tb]
\centering
\includegraphics[width=0.9\textwidth]{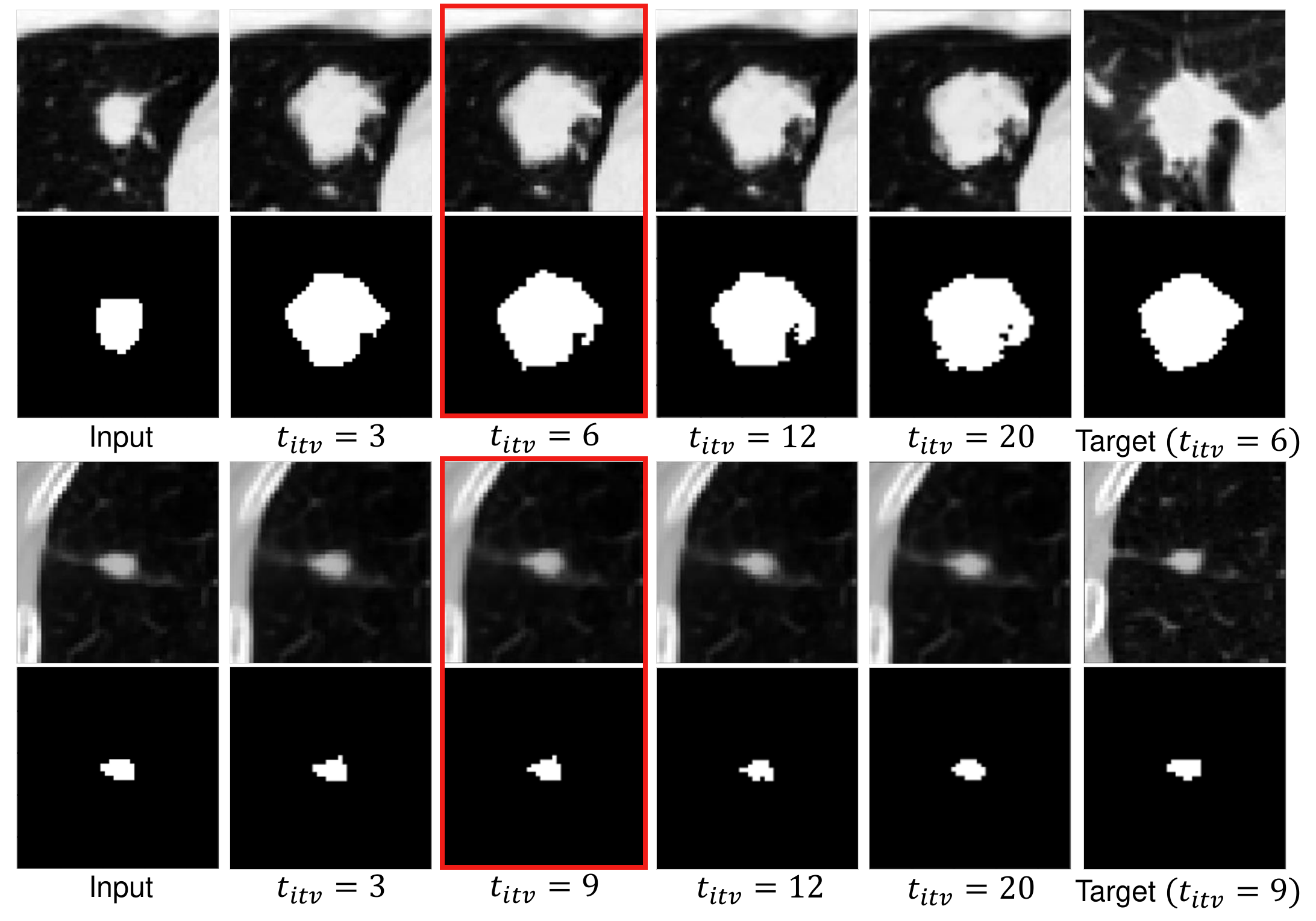}
\caption{Continuous prediction results of a PD case (top) and a non-PD case (bottom) by WarpNet. The first and last columns are the input image/segmentation and the target image/segmentation, and columns in the middle are the warped images/segmentations by WarpNets with different temporal encodings. Warped results that have the same time interval as the targets are highlighted in red.} \label{fig:timeline}
\end{figure}

Our \textit{NoFoNet} is trained to predict what the nodule may be visually like after a certain time interval, then we can determine whether it is a PD (progressive disease, i.e., significant growth of nodule in size) case. Since some nodules in our dataset have multiple follow-ups, we stipulate that a nodule is judged as a PD case as long as one of its follow-up pairs (see Sec. 2.1) meets specific criterion, which is determined with the help of two senior radiologists.

Define $V_1$, $V_2$ ($\mathrm{mm}^3$) as the two nodule volumes of a follow-up pair with time interval $T$ ($\mathrm{d}$), the criterion is as follows: (1) Considering that the fast-growing nodules have higher risks, we calculate the average volume growth rate $\mathrm{AVGR} = (V_1 - V_2) / T$, and set a threshold of 1; (2) Some cases may have $\mathrm{AVGR}$ less than 1 but eventually grow significantly in size, we set a threshold of 200 for volume difference $\mathrm{VD} = V_1 - V_2$ and a threshold of $50\%$ for relative volume difference $\mathrm{RVD} = (V_1 - V_2) / V_1$. In summary, a nodule is classified as PD case if one of its observed $\mathrm{AVGR}\geq1\mathrm{mm}^3/\mathrm{d}$, or $\mathrm{VD}\geq200\mathrm{mm}^3$ and $\mathrm{RVD}\geq50\%$.

The 315 nodules are divided into two parts according to the aforementioned criterion, resulting in 64 positive cases (PD, significant growth) and 251 negative cases (non-PD, stable or shrinking).
We split our dataset randomly into 5 groups based on patients (i.e., all nodules of one patient must be in same subset) and perform 5-fold cross validation to evaluate our models.


\subsection{Performance Analysis}

In this section we will present some quantitative results and qualitative results. Table~\ref{tab1} shows the performance of our models and baselines using 5-fold cross validation method. Note that U-Net w/ or w/o TEM predicts output images directly so it only has PSNR and $\mathrm{PSNR}^*$ (PSNR in the nodule parts) for output/target images. As is shown in Appendix Fig. A.1, U-Net baselines generate predicted images with low visual quality. It is noticeable that when added segmentation loss for warped/target images, WarpNet predicts more accurate displacement fields, resulting in higher dice coefficient between warped/target images and better performance for PD/non-PD classification than WarpNet without segmentation loss. We use the geometrical mean (G-mean) of sensitivity (TP/TP+FN) and specificity (TN/TN+FP) as main evaluation index for the unbalanced dataset. The TextureNet in \textit{NoFoNet} improves the visual quality of the warped images and achieves higher PSNR/$\mathrm{PSNR}^*$ scores, as visually shown in Fig.~\ref{fig:warp-texture}.

Fig.~\ref{fig:warp-texture} shows the results of spatial transformation for input images by WarpNet and voxel-wise texture addition for warped images by TextureNet. We select a PD case and a non-PD case to demonstrate the performance of \textit{NoFoNet} on different types of nodules. It can be seen that TextureNet is able to refine the warped images from WarpNet and increase the intensity similarity between the predicted and target nodules. Please refer to Appendix Fig. A.1 for more comparison results (including results from U-Net).

Fig.~\ref{fig:timeline} illustrates the continuous prediction results of two nodules using WarpNet. Note that results with the same follow-up time interval as the targets are highlighted in red. We choose a PD (progressive disease) case and a non-PD case for contrast to show that our WarpNet can represent both significant growth and stabilization of nodules in size well. For PD case it can also be seen that the model is able to generate reasonable nodules as time interval changes and the variation tendency is plausible, indicating the effectiveness of TEM.

\section{Conclusion}
We develop the \textit{NoFoNet}, a unified network to predict the tumor growth for lung nodules. By explicitly learning spatial transformation and texture transformation, it yields high-quality visual appearances and accurate quantitative results, with validated effectiveness on an in-house dataset from two clinical centers. To the best of our knowledge, this is one of the first study to predict nodule growth quantitatively (size) and visually (appearance) given any time interval during.

A limitation of this study is that we only model the tumor growth as the indicator of nodule risk. However, according to TNM tumor staging system, tumor size (T), lymph node (N) and metastasis (M) are considered in tumor prognosis assessment. In future studies, we will address the N and M information to develop a more advanced risk stratification system for lung nodule follow-up. Besides, we will expand the dataset from cooperative hospitals and explore more effective architectures and temporal encoding methods for our framework.

\subsubsection{Acknowledgment.}
This work was supported in part by National Natural Science Foundation of China (61671298), 111 project (BP0719010), Shanghai Science and Technology Committee (18DZ2270700) and Shanghai Jiao Tong University Science and Technology Innovation Special Fund (ZH2018ZDA17). The corresponding author of this paper is Yi Xu (xuyi@sjtu.edu.cn).

%
%
%
%


\bibliographystyle{splncs04}







\end{document}